\documentclass[prl,aps,twocolumn,floatfix,nofootinbib]{revtex4}
\usepackage{epsf,color,colordvi,amsmath,epsfig,epstopdf}
\usepackage{epsfig}
\usepackage{feynmf}		
\usepackage{graphics}
\usepackage{subfigure}

\def\inbar{\,\vrule height1.5ex width.4pt depth0pt}
\def\IR{\relax{\rm I\kern-.18em R}}
\def\IC{\relax\hbox{$\inbar\kern-.3em{\rm C}$}}

\def\weakangle{\sin^2{\theta}_W}

\newcommand{\isotope}[2]{$^{#2}{\rm #1}$}

\begin{document}

\title{Sterile Neutrinos, Coherent Scattering and Oscillometry Measurements with Low-temperature Bolometers}

\author{Joseph A. Formaggio}
\author{E. Figueroa-Feliciano}
\author{A.J. Anderson}
\affiliation{Massachusetts Institute of Technology,\\
Cambridge, MA 02139}

\begin{abstract}  
Coherent neutrino-nucleon scattering offers a unique approach in the search for physics beyond the Standard Model.  When used in conjunction with mono-energetic neutrino sources, the technique can be  sensitive to the existence of light sterile neutrinos.  The ability to utilize such reactions has been limited in the past due to the extremely low energy threshold ($10-50$ eV) needed  for detection.  In this paper, we discuss an optimization of cryogenic solid state bolometers that enables reaching extremely low kinetic energy thresholds.  We investigate the sensitivity of an array of such detectors to neutrino oscillations to sterile states.  A recent analysis of available reactor data appears to favor the existence of such such a sterile neutrino with a mass splitting of $|\Delta m_{\rm sterile}|^2 \ge 1.5$ eV$^2$ and mixing strength of $\sin^2{2\theta_{\rm sterile}} = 0.17\pm 0.08$ at 95\% C.L.  An array of such low-threshold detectors would be able to make a definitive statement as to the validity of the interpretation.
\end{abstract}                                                                 

\date{\today}
\maketitle

\section{Introduction}
\label{sec:intro}

The theory of neutrino oscillations, as described by the Pontecorvo-Maki-Nakagawa-Sakata (PMNS) mixing matrix, provides a simple and well-grounded description of the neutrino data obtained thus far~\cite{Nakamura:2010zzi}.  Neutrino oscillation experiments carried out over the past half-century firmly establish the presence of the phenomena to the point that the existence of non-zero neutrino masses is no longer in question.  Current experiments are now engaged in gaining greater precision of the relevant mixing and mass parameters in a manner similar to that which was done for the quark sector.

As the precision of such experiments continues to improve and the tools by which the data is analyzed increase in sophistication, one finds that the emerging picture may be more complex than previously realized.  For example, a recent re-analysis of existing reactor data by Mention and collaborators~\cite{Mention:2011rk} shows a $\simeq 3 \sigma$ deviation from theoretical predictions.  Though it is possible that the discrepancy could be due to difficult to calculate Standard Model effects, such as weak magnetism~\cite{Huber:2011wv}, the result also appears consistent with the presence of a fourth, sterile neutrino.  A combined analysis using available reactor data, as well as data collected by gallium solar neutrino calibration experiments~\cite{Anselmann:1995ag,Abdurashitov:2005tb} and the MiniBooNE neutrino data~\cite{AguilarArevalo:2007it} leads to a mass splitting of $|\Delta m_{\rm sterile}^2| > 1.5$ eV$^2$ and $\sin{(2\theta_{\rm s})}^2 = 0.17 \pm 0.08$ at 95\% C.L.  This observation appears to be further corroborated by other measurements, particularly the more recent data collected by the MiniBooNE experiment in antineutrino running~\cite{AguilarArevalo:2010wv}.   At this stage it is too early to make any strong claim as to the validity of one or all of these observations.  Continued data collection and scrutiny of systematic uncertainties will provide better guidance as to whether new physics is at play.

The most likely beyond the Standard Model explanation given to the observations made at LSND and MiniBooNE and, more recently, from reanalysis of reactor data is the existence of at least one sterile neutrino with a mass scale of 1 eV.  Ongoing short-baseline measurements, as well as other complimentary approaches~\cite{bib:Formaggio2011}, should be able to determine if the data continues to diverge from Standard Model predictions.  However, to lay claim that the observation is indeed that of a sterile neutrino would almost certainly warrant one or more experiments with unique signatures to the phenomena.  In this paper, we propose an alternate approach which makes use of oscillometry measurements of neutrino-nucleon coherent scattering in order to positively confirm or refute the existence of sterile neutrinos.

\section{Detection via Coherent Scattering}

Coherent scattering offers distinct advantages compared to other techniques in disentangling the signature of sterile neutrinos.  First and foremost, coherent scattering off nuclei is a neutral current process.  Thus, any observation of an oscillation structure would indicate mixing solely to non-active neutrinos.  Other methods, such as neutrino-electron scattering, must disentangle the mixing to sterile neutrinos from mixing to active neutrinos.  The technique becomes even more powerful when combined with low energy mono-energetic sources.  Oscillations to neutrinos at the eV mass scale would manifest themselves over the length of a few meters (for $\sim 1$ MeV neutrino energies).  The signature would be quite difficult to mimic with typical backgrounds.  Finally, the cross-section for the process is greatly enhanced thanks to the coherent nature of the reaction.

The use of intense neutrino sources to probe sterile neutrinos has been proposed previously in the literature~\cite{Vergados:2011gi,Vergados:2010sj,Agarwalla:2010gd,PhysRevD.75.093006}.  The difficulty with all such detection schemes is the low energy threshold necessary to detect the signature nuclear recoil.  Such difficulties are circumvented by either resorting to targets with low mass numbers--considerably lowering the cross-section amplitude and requiring large mass detectors--or by looking instead at the charged current reaction using higher energy neutrinos.   In this paper, we discuss a low energy threshold detector based on cryogenic bolometers that has the capability of reaching recoil energy thresholds as low as 10~eV.  Such detectors re-open the door to neutral current coherent scattering as a method for sterile neutrino detection.

Neutrino-nucleus interactions which are coherent in character have the advantage of scaling as $A^2$, where $A$ is the mass number of the target nucleus.  For a target nucleus with atomic number $Z$ and neutron number $N$, the cross-section as a function of recoil kinetic energy is given by the expression~\cite{bib:Freedman}:

\begin{equation}
\frac{d\sigma(\nu A \rightarrow \nu A)}{dT}  = \frac{G_F^2}{4\pi} M_A Q_W^2 (1 - \frac{M_A T}{2 E_\nu^2}) F(q^2)^2
\end{equation}

\noindent where $G_F$ is the Fermi coupling constant, $M_A$ is the mass of the nucleus, $F(q^2)$ is the nuclear form factor, and $Q_W$ is the weak charge, defined by the relation:

\begin{equation}
Q_W = N - Z(1-4\weakangle)
\end{equation}

In our study, we will mainly consider mono-energetic electron capture sources, all of which have neutrino energies below 1 MeV.  The maximum momentum transfer for such sources is $|q_{\rm max}| \le 2 E_\nu \ll 2$ MeV.  Since the form factor $F(q^2) \rightarrow 1$ for cases where the scale of the momentum probe is much larger than the size of the nucleus, we can safely ignore this correction factor for our analysis.

The maximum kinetic energy imparted on the nuclear recoil depends on the neutrino energy and the mass of the recoil target:

\begin{equation}
T_{\rm max} \le \frac{E_\nu}{1+\frac{M_A}{2E_\nu}}
\end{equation}


For a silicon target at 1 MeV, that implies a maximum kinetic energy of about 50 eV.  For a germanium target the maximum kinetic energy would be around 20 eV. Such low kinetic energies are why detection of the process has been so elusive to date.  The fraction of events that is detectable by a given experiment depends crucially on the inherent threshold of the detector.  For a monochromatic source of energy $E_\nu$, the effective cross-section can be written as:

\begin{eqnarray}
\bar{\sigma} = \int_{T_0}^{T_{\rm max}} \frac{d \sigma}{dT}(E_\nu) \cdot dT \\
\bar{\sigma} =  \sigma_0(E_\nu) \cdot f(E_\nu, T_0)
\end{eqnarray}

\noindent where $\sigma_0(E_\nu) \equiv \frac{G_F^2}{4\pi} E_\nu^2 Q_W^2 $ is the total integrated cross-section assuming no energy threshold and $f(E_\nu, T_0)$ represents the fraction of events above a given threshold energy, $T_0$.  In the limit that $E_\nu \ll M_A$, the fraction of events above threshold can be written as:

\begin{equation}\label{eqn:fracAboveThresh}
f(E_\nu, T_0) = (1-\frac{T_0}{T_{\rm max}})^2
\end{equation}

Any detector hoping to detect such a signal with sufficient statistics must achieve as low a recoil threshold as possible.

\section{The \isotope{Ar}{37}  Source}
\label{sec:source}

Oscillometry-based measurements benefit greatly from the use of mono-energetic neutrino sources, since it reduces the measurement to a pure flux-versus-distance analysis.  Low energy electron capture sources provide the most effective and clean source of such neutrinos available to date~\cite{Haxton:1988ee}.  A number of such neutrino sources have been considered in the literature; a few of them are listed in Table~\ref{tab:sources}.  Historically, two such high intensity source have been produced for neutrino studies:  a \isotope{Cr}{51} source, used by the SAGE and GALLEX experiments~\cite{PhysRevC.59.2246,Anselmann:1994ar}, and an \isotope{Ar}{37} gaseous source used in conjunction with the SAGE experiment~\cite{Abdurashitov:2005tb}. 

\begin{table*}[htdp]
\caption{List of properties of selected electron capture neutrino sources.}
\begin{center}
\begin{tabular}{|c|c|c|c|c|c|}
\hline
Source & Half-Life & Progeny & Production  & $E_\nu$ & Gamma (?) \\
\hline
\isotope{Ar}{37} & 35.04 days & \isotope{Cl}{37} &  \isotope{Ca}{40}(n,$\alpha$)\isotope{Ar}{37}  & 811 keV (90.2\%), 813 keV (9.8\%) & Inner Brem only \\
\isotope{Cr}{51} & 27.70 days & \isotope{V}{51} & n capture on \isotope{Cr}{50}& 747 keV (81.6\%), 427 keV (9\%), 752 keV (8.5\%) & 320 keV $\gamma$ \\
\isotope{Zn}{65} & 244 days & \isotope{Cu}{65} & n capture on \isotope{Zn}{64} & 1343 keV (49.3\%), 227 keV (50.7\%)& 1.1 MeV $\gamma$ \\
\hline
\end{tabular}
\end{center}
\label{tab:sources}
\end{table*}%

The \isotope{Ar}{37} source is perhaps the most ideal with respect to a future coherent-scattering measurement, for a number of reasons:

\begin{itemize}
\item \isotope{Ar}{37} produces a very high-energy, near mono-energetic neutrino (90.2\% at 811 keV, 9.8\% at 813 keV).
\item With the exception of inner bremsstrahlung photons, almost all the energy is carried away by neutrinos, facilitating shielding and enabling the source to be extremely compact.
\item Extremely high production yield per reactor target.
\end{itemize}

The SAGE collaboration successfully produced such a source with a total activity of about 400 kCi to be used in conjunction with their gallium solar neutrino detector.  The source was also very compact, extending 14 cm in length and 8 cm in diameter, including shielding~\cite{bib:Ar37}.  Further reduction in size might be possible, even with increased activity, making \isotope{Ar}{37} an ideal portable neutrino source.  

Despite its clear advantages as a source and its historical precedent, production of such sources is less than ideal.  The reaction process by which it is generated (\isotope{Ca}{40}(n,$\alpha$)\isotope{Ar}{37}) requires a high fast neutron flux above 2 MeV, an energy regime where few reactors operate~\cite{Michael1984813,bib:Bruce}.  Production also requires large amounts of CaO and processing in nitric oxide, which makes post-production handling difficult.  Far less complex to produce is \isotope{Cr}{51}, which requires only thermal neutrons capturing on \isotope{Cr}{50}.  However, as a source, the high energy gamma produced from the decay of the excited state of \isotope{V}{51} imposes more shielding requirements.  As such, intense  \isotope{Cr}{51} may be less ideal for this investigation, but still worth considering given the advantages in producing the required activity.  

With it's high energy neutrino emission, \isotope{Zn}{65} is also an attractive source for consideration~\cite{bib:Zn65}.  However, its 1.1 MeV gamma emission complicates the shielding, so this source is not considered further.

\section{The Detector}
\label{sec:detector}

The detector requirements for this experiment are extremely challenging.  Due to the low energy of the neutrinos ($\le1$~MeV), the recoil energy deposited in the target is in the order of tens of eV, while the minimum mass needed is hundreds of kilograms. Methods of determining the energy deposition from particle interactions in a target include measuring the ionization, the scintillation, and/or the phonon excitations in the material. For nuclear recoils of tens of eV, the fraction of the energy deposited by the scattering event that produces free or conduction band electrons (the quenching factor) is unknown at these energies, and is expected to be very low (could be zero for some materials). Thus any readout scheme involving ionization channels will be at a severe disadvantage. Similar uncertainties hold for the scintillation yield from nuclear recoils at these energies. An additional problem for both ionization and scintillation readout is that the energy required to create a single electron, electron-hole pair, or scintillation photon from a nuclear recoil in most liquid or solid targets is a few eV for ionization and tens of eV for scintillation. Thus, even if any quanta were produced, Poisson statistics would make the measurement of the energy of any given recoil event fairly poor. We have therefore focused our attention to the measurement of phonons created in the interaction. With mean energies of the order of $\mu$eV, thermal phonons provide high statistics at 10 eV and sample the full energy of the recoil with no quenching effects.

\subsection{Historical Precedent}
Bolometric detection of neutrino interactions was one of the prime drivers for the development of the low-temperature detector community. The idea of searching for signs of coherent (or ``gentle'') neutrino scattering with cryogenic bolometers was first suggested by Lubkin~\cite{Lubkin:1974hs}, quickly followed with the first experiment design by Niinikoski and Udo~\cite{Niinikoski:1974te} for detecting coherent scattering of neutrinos from an accelerator or reactor using 1~cm$^{3}$ silicon bolometers at 5~mK with an estimated energy resolution of 2~eV. Their model did not incorporate the heat capacity of the thermometer or the thermal coupling of the thermometer to the target, but established a low-energy threshold very similar to what we propose in this paper.

Cabrera, Krauss and Wilczek~\cite{Cabrera:1985gn} proposed a multi-ton silicon bolometer array using deposited superconducting films as a thermometer to detect neutrinos from reactors and the sun. The desire for large total masses and the higher available neutrino energies from these sources pushed the detector optimization to an array of kg-scale silicon targets, with thermodynamic-noise-limited thresholds in the hundreds of eV.

Rare event searches with cryogenic crystal bolometers are being actively pursued by several groups in neutrino physics~\cite{CUORE,CUORICINO,MARE} and dark matter searches~\cite{CRESST,CDMS,EDELWEISS,ROSEBUD}. All of these experiments have energy thresholds hundreds or thousands of times higher than the desired threshold for this experiment. This is due to optimizations of the science reach for a given experimental low energy threshold, operating temperature, detector mass, and readout technology.

The maximum recoil energy induced in the silicon target by our proposed neutrino source is around 50~eV. To achieve the required sensitivity, we have taken the approach suggested by Niinikoski of using arrays of small Si bolometers. We chose, however, to follow the approach in~\cite{Cabrera:1985gn} and use transition-edge sensors (TES) as the temperature readout. We took these ideas and optimized the design for this sterile neutrino search. 

\subsection{Detector Design and Expected Performance}

The threshold for a bolometer is a function of its baseline energy resolution. A dimensionless measure of the sensitivity of a resistive thermometer at a temperature $T$ and resistance $R$ is the quantity $\alpha$, defined as $\alpha \equiv \frac{T}{R}\frac{dR}{dT}$. The energy resolution of a TES bolometer is approximately~\cite{Irwin:1995ie}
\begin{equation}
\Delta E_{\rm rms} = \sigma_{E} \approx \sqrt{\frac{4 k_{B}T^{2} C_{\rm tot}}{\alpha} \sqrt{\frac{\beta+1}{2}}},
\end{equation}
where $k_{B}$ is the Boltzmann constant, $C_{\rm tot}$ is the total heat capacity of the bolometer, and $\beta$ is the exponent of the temperature dependance of the thermal conductivity between the bolometer and the refrigerator. To unambiguously detect events above the noise from the detector, we set the experimental threshold to 7.5 $\sigma_{E}$. For a 10~eV threshold, we then need a detector with $\sigma_{E} < 1.33$~eV, or expressed in terms of the full width at half maximum, $\Delta E_{\rm FWHM} = 2\sqrt{2 \ln 2} \, \sigma_{E} < 3.14$~eV.

Assuming a conduction path to the cold bath of the refrigerator dominated by Kapitza resistance, $\beta = 4$, and with a temperature $T=15$~mK, a 10~eV threshold could be attained with a heat capacity $C_{\rm tot} \leq 200$~pJ/K.  However, this model is not complete, as it assumes a perfectly isothermal bolometer. In practice, the various internal heat capacity systems of the bolometer are decoupled from each other through internal conductances, and thermalization times of each separate heat capacity must also be taken into account. These internal decouplings introduce various sources of noise, degrading the energy resolution of the bolometer and consequently requiring a smaller heat capacity to attain the desired threshold. 

\begin{figure}[htbp]
\begin{center}
\includegraphics[width=0.8\columnwidth,keepaspectratio=true]{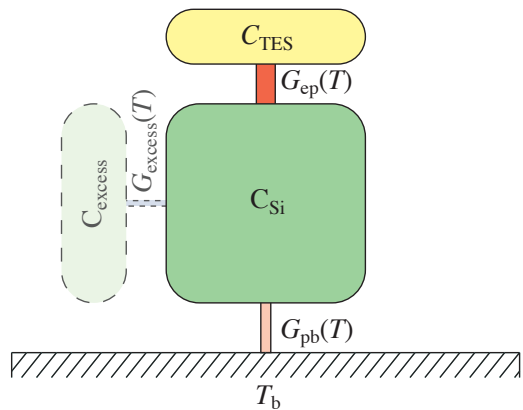} 
\caption{Schematic of bolometer model. The refrigerator acts as a cold bath at temperature $T_{b}$. The Si heat capacity is connected by a thermal conductance $G_{\rm pb}$ to the bath. The TES is connected by the electron-phonon conductance $G_{\rm ep}$ to the Si. A potential excess heat capacity with its coupling are shown in dashed outlines. For this study we have assumed $C_{\rm excess}$ and/or $G_{\rm excess}$ can be made small enough to become negligible.}
\label{fig:complexcal}
\end{center}
\end{figure}

Fig.~\ref{fig:complexcal} shows a schematic of the model.  The bolometer is connected to the cold bath at temperature $T_{\rm b}$ through a weak thermal conductance $G_{\rm pb}$. The total heat capacity can be described by $C_{\rm tot} = C_{\rm Si} + C_{\rm TES} + C_{\rm excess}$, where $C_{\rm Si} \propto T^{3}$ is the theoretical heat capacity of Si given by Debye theory, $C_{\rm TES} \propto T$ is the TES heat capacity dominated by the metal electron system, and $C_{\rm excess}$ is the heat capacity of impurity bands and two-level systems in the crystal.  The TES phonon system is assumed to be at the same temperature as the silicon phonon system, since the sub-micron thickness of the TES makes it incapable of sustaining its own thermal phonon population. The TES electron system is coupled to the phonon system through its electron-phonon coupling conductance $G_{\rm ep}$.

Impurities and defects in the Si substrate can lead to impurity bands and two-level systems (TLS) in the crystal which add to the total heat capacity of the system. The size of this excess heat capacity and its equilibration time with the phonon system depends on the specific mechanism involved~\cite{Keyes:1988}. Si and Ge crystals can be acquired with impurity levels of $\sim 10^{15}$ and $10^{10}$~atoms/cm$^{3}$~\cite{Armengaud:2010fn}, respectively. At these levels the heat capacity $C_{\rm excess}$ and/or the thermal conductance $G_{\rm excess}$ could be low enough to render them negligible for our purposes. Ge clearly holds a large advantage in this regard, but its lower maximum recoil energy for coherent neutrino scatters and higher Debye temperature make the design of the detector more challenging, thus making it our backup if the Debye heat capacity values cannot be achieved in Silicon at 15 mK.

Knaak and Mei{\ss}ner~\cite{Knaak:1984wb} measured the heat capacity of a high-purity 19~g silicon crystal with the following impurity concentrations: B: $\sim5\times10^{12}\, \rm atoms/cm^{3}$,  O and C: $\sim1\times10^{15}\, \rm atoms/cm^{3}$. Their measurement was done down to 50~mK temperatures, very similar in mass, impurity concentration and temperature to what we wish to use in our detector. They measure the time-dependent heat capacity on three time scales, initial ($< 1$~ms), short ($\sim 1$~ms), and long ($> 0.1$~s). The initial time scale measures the athermal signal before thermalization and is less relevant to our present discussion. For the short time constant, they report a measured heat capacity consistent with the Debye prediction. For the long time constant, they see a larger heat capacity which they attribute to the heat capacity of the gold, carbon, and epoxy used to instrument their sample. It is possible that some of the excess heat capacity in their measurement was not due to the instrumentation attached to the sample. The time constant of our proposed detector design is 50~ms, in between these two reported measurement scales. More measurements should be done at these low temperatures to study the heat capacity dependence of Si with different species and concentrations of impurities. 

The Knaak and Mei{\ss}ner results suggest that low heat capacities are achievable. In our design, we benefit from the fact that our TES thermometer makes up about half of the total heat capacity, which allows us to tolerate some excess heat capacity from the silicon target. We also have the option to optimize using less mass per bolometer, and trading off overall experiment mass for lower threshold per detector. 

For this study, we will assume the excess heat capacity is negligible, and optimize the heat capacity of the thermometer $C_{\rm TES}$, the electron-phonon coupling $G_{\rm ep}$, and the Si heat capacity $C_{\rm Si}$ to obtain the desired 10~eV threshold with the highest possible target mass. We make the following assumptions:

\begin{itemize}
\item Each detector is a Si cube ranging in mass from 20--100~g. The heat capacity is determined from Debye theory.
\item The conductance between the Si and the cold bath, $G_{\rm pb}$, can be engineered to give a desired value. The value is chosen to give a thermal impulse response time of 50~ms as measured by the thermometer readout. 
\item The thermometer is a Mo/Au TES bilayer with a superconducting transition engineered to a specific temperature between 10--100~mK. Mo/Au TES X-ray detectors have achieved resolutions of $\Delta E_{\rm FWHM}=2$~eV~\cite{Bandler:2008fv}.
\item The TES heat capacity and electron-phonon coupling are taken from the literature and are a function of the chosen volume of the TES and the temperature.
\end{itemize}

Given these general assumptions, several combinations of detector mass and transition temperature were tested for both Si and Ge targets, scaling the TES volume to obtain the best energy resolution, following the theoretical framework of~\cite{FigueroaFeliciano:2006bc}. The TES volume is a compromise between two competing interests: having a small TES heat capacity, and having a fast thermal link between the TES and the Si or Ge target. The optimum volume corresponds to a TES heat capacity that is roughly equal to the target.  An important quantity is the ratio $G_{\rm ep}/G_{\rm pb}$. As long as this ratio is $\gtrsim 100$, the TES  remains in quasi-thermal equilibrium with the target throughout a pulse (except for the initial athermal phase on the order of 1~ms). If needed, one can make $G_{\rm pb}$ smaller and gain energy resolution (and lower threshold) at the expense of slower signals. 

\begin{figure}[htbp]
\begin{center}
\includegraphics[width=1.\columnwidth,keepaspectratio=true]{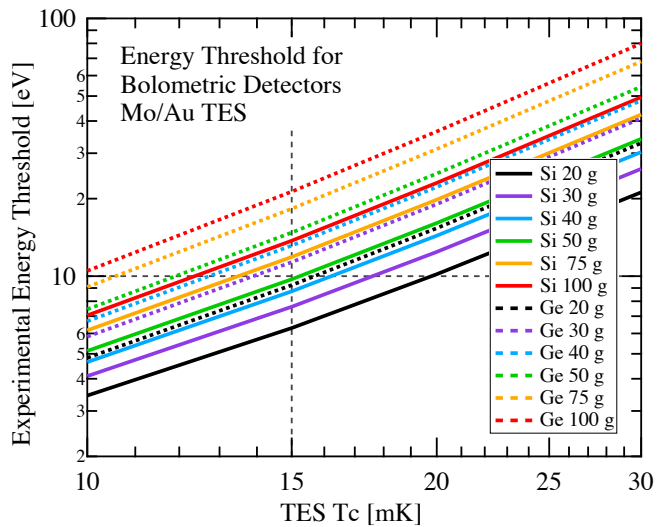} 
\caption{Calculation of threshold for Si and Ge targets from 20--100~g at different operating temperatures. The lowest line for each target material is the 20~g line. The model (see Fig~\ref{fig:complexcal}) takes into account the heat capacity of the TES and the target, the internal thermal fluctuation noise between the target and the TES thermometer, the electronics noise, the Johnson noise from the TES and its bias resistor, and the phonon noise between the target to the bath. The volume of the TES was scaled to give the best energy resolution at 15~mK. The horizontal dashed line marks the desired 10~eV threshold, corresponding to an energy resolution $\Delta E_{\rm FWHM}=3.14$~eV. The vertical dashed line marks the desired operating temperature of 15~mK. For Si, a 50~g target meets the requirements. For Ge, a 20~g target meets the requirements.}
\label{fig:thresh}
\end{center}
\end{figure}

\begin{figure}[htbp]
\begin{center}
\includegraphics[width=1.\columnwidth,keepaspectratio=true]{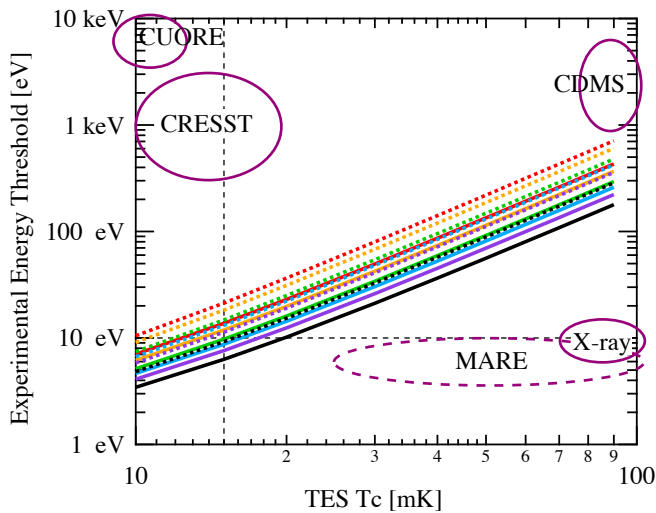} 
\caption{Zoom-out of Fig.~\ref{fig:thresh}. Our desired operating point is designated by the vertical and horizontal dashed lines. The legend for both plots are the same. The ovals designate the region of operation for the following low-temperature experiments: CUORE~\cite{CUORICINO}, EDELWEISS~\cite{Armengaud:2010fn}, CRESST~\cite{Lang:2010bt}, CDMS~\cite{CDMS}, X-ray microcalorimeters for astrophysics~\cite{Bandler:2008fv}, and MARE (proposed)~\cite{MARE}. Note that these experiments use different combinations of thermal or athermal measurements, TES or NTD thermometry, and mass per detector unit, so cannot be directly compared to the design presented in this paper.}
\label{fig:threshzoom}
\end{center}
\end{figure}

The results of our models are shown in Fig.~\ref{fig:thresh}. The plotted threshold is calculated as $7.5~\sigma_{E}$. Due to practical limitations in refrigeration and considering the readout necessary for the size of the experiment, we focus on a transition temperature of 15~mK, with the refrigerator base temperature at 7.5~mK. At this temperature, a low-energy threshold of 10~eV can be obtained with bolometers with 50~g of Si or 20~g of Ge. A 50~g Si target sees about twice the rate of neutrino coherent recoil events as a 20~g Ge target when both have a 10~eV threshold. We will thus focus on Si. Model parameters for the Si detector are given in Table~\ref{tab:Sipars}. 
\begin{table}[htdp]
\caption{Model parameters for a 50~g Si target coupled to a Mo/Au TES operated at 15~mK. The Si target is a 28~mm cube, and the TES is an 25~mm $\times$ 2~mm film 600~nm thick deposited on the Si surface. The energy resolution for this model is 3~eV FWHM, with a 10~eV threshold. Pulses from this model are shown in Fig~\ref{fig:pulses}.}
\begin{center}
\begin{tabular}{|l|c|c|l|}
\hline
Parameter & Value & Units & Description \\
\hline
$C_{\rm Si}$ 	& 43.3	& pJ/K	& Debye heat capacity\\
$C_{\rm TES}$	& 31.1	& pJ/K	& TES electron heat capacity\\
$G_{\rm ep}$	& 29.3	& nW/K	& TES-Si thermal conductance\\
$G_{\rm pb}$	& 0.17	& nW/K	& Si-bath thermal conductance\\
$T_{b}$		& 7.5		& mK	&Cold bath temperature\\
$T_{c}$		& 15		& mK	&TES temperature\\
$R_{o}$		& 3		& m$\Omega$ & Quiescent TES resistance\\
$I_{o}$		& 14.1	& $\mu$A	& Quiescent TES current\\
$P_{o}$		& 0.6		& pW	& Quiescent TES power\\
$\alpha=\frac{T_{c}}{R_{o}}\frac{dR}{dT}$		& 50		& - & TES sensitivity\\
$\tau_{\rm o}$	& 436.2	& ms		& Natural decay time $C_{\rm tot}/ G_{\rm pb}$\\
$\tau_{\rm eff}$	& 51.1	& ms		& Response time with TES speedup\\
$\tau_{\rm decay}$	& 29.2	& ms		& Decay time with readout circuit\\
$L$			& 30		& $\mu$H & Readout inductance\\
\hline
\end{tabular}
\end{center}
\label{tab:Sipars}
\end{table}%

\begin{figure}[htbp]
\begin{center}
\includegraphics[width=1.\columnwidth,keepaspectratio=true]{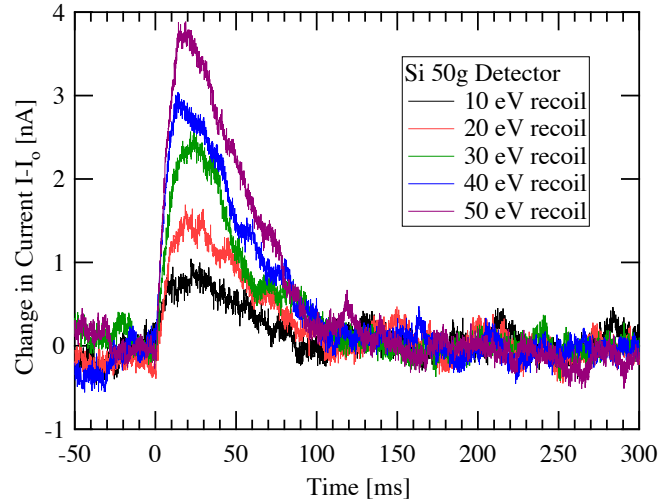} 
\caption{Simulated current readout for 10--50~eV recoils using the model parameters in Table~\ref{tab:Sipars}. The current has been multiplied by -1 to make the pulses positive. Noise sources modeled are: the phonon noise between the target to the bath, the internal thermal fluctuation noise between the target and the TES thermometer, the Johnson noise from the TES and its bias resistor, and the electronics noise. The modeled 10~eV pulses are clearly separated from the noise.}
\label{fig:pulses}
\end{center}
\end{figure}

The natural decay time of the bolometer $C_{\rm tot}/G_{\rm pb}=436$~ms. Electro-thermal feedback~\cite{Irwin:1995ie} from the TES speeds up the response time to roughly 50~ms. The TES readout is bandwidth-limited by an inductor which critically damps the system, causing a further speedup in the response. The decay time of recoil events is reduced from the 50~ms decay with no inductor to a 30~ms decay with the inductor. 
Fig.~\ref{fig:pulses} shows a simulation of 10--50~eV neutrino coherent scatters in a 50~g Si bolometer. The pulses are clearly separated from the noise, and the energies of the different events are clearly separated by eye. 

The total heat capacity is on the order of 460~keV/mK, and given a transition width of around 1~mK for a TES, we estimate that the bolometer will have fairly linear response up to hundreds of keV. Higher energies will have a non-linear response but will retain significant energy resolution. This large energy bandwidth will help understand the background in our experiments, and enable other rare event searches such as limits on the neutrino magnetic moment and dark matter interactions.

Multiplexing readout schemes for transition-edge sensors are now a mature technology being developed for many astronomical applications, for example~\cite{Reintsema:2009er}, and 10,000 channel systems with time constants similar to this application are already in operation~\cite{Bintley:gz}. Schemes for even larger multiplexing gains are in development~\cite{Niemack:2010gs}. Given the slow time constants of this application, a 10,000 channel multiplexer design carries a fairly low risk and would allow 500~kg of Si to be instrumented.

\begin{figure}[htbp]
\begin{center}
\includegraphics[width=0.8\columnwidth,keepaspectratio=true]{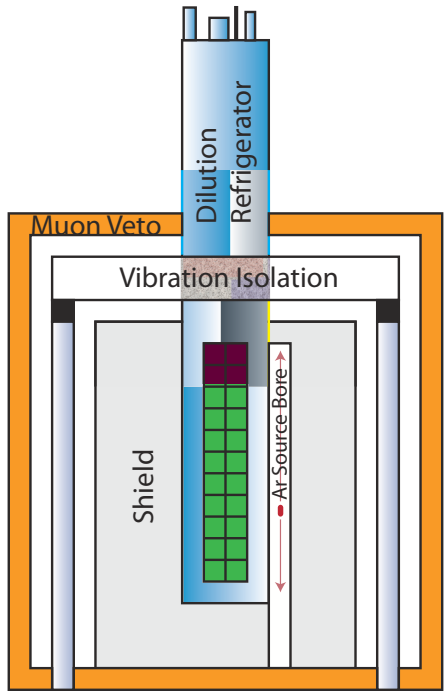} 
\caption{Conceptual schematic of the experimental setup for a bolometric measurement of coherent scattering from a high-intensity \isotope{Ar}{37} neutrino source.  An array of 10,000 Si bolometers is arranged in a column of dimensions  0.42 (dia.)~$\times$~2.0~(length)~meters (shown in green) inside a dilution refrigerator suspended from a vibration isolation mount.  Each Si bolometer has a mass of 50 g for a total active mass of 500~kg. Appropriate passive or active shielding surrounds the refrigerator. A cylindrical bore in the shield allows the \isotope{Ar}{37} source, mounted on a translation mechanism, to be moved to different positions along the side of the array. Periodic movement of the source throughout the measurement sequence allows each detector to sample multiple baselines, enables cross-calibration among detectors, and aids in background subtraction. The minimum distance from the source to a bolometer is assumed to be $\sim$10~cm.}
\label{fig:array}
\end{center}
\end{figure}

A concept for a 500~kg payload is shown in Fig.~\ref{fig:array}. The 10,000 Si bolometers are arranged in a column of dimensions 0.42 (dia.)~$\times$~2.0~(length)~meters inside a dilution refrigerator suspended from a vibration isolation mount.  Passive or active shielding surrounds the refrigerator. The exact shape and type of shielding will be determined at a later time. A cylindrical bore, perhaps 10~cm or less in diameter, is removed from the shield and allows the \isotope{Ar}{37} source, mounted on a radio-pure translation mechanism, to be moved to different positions along the side of the array. Periodic movement of the source throughout the measurement sequence allows each detector to sample multiple baselines, enables cross-calibration among detectors, and aids in background subtraction. The minimum distance from the source to a bolometer is assumed to be $\sim$10~cm.

\subsection{Detector Backgrounds}

The detectors described in the previous section will be sensitive to several sources of background in the recoil energy range of 10-50~eV.  Unfortunately, it is difficult to estimate accurately the rate of events from each of these sources, and the levels expected are currently unknown.  Although we do not have a quantitative understanding of the backgrounds in this regime, it is important to note that backgrounds can be measured and subtracted using data taken when the neutrino source is not in place.  We qualitatively consider several sources of background which we expect to be present in the energy range 10-50~eV:

\begin{itemize}
\item {\it Radiogenic impurities:} Most radioactive decay products have energies in the range of hundreds of keV to tens of MeV and will be clearly distinguishable from the neutrino signal.  Radiogenic impurities may still contribute to the background in two primary ways.  First, many common impurities produce gamma rays that can interact with material by the photoelectric effect or Compton scattering and produce background events by the mechanisms described below.  These gammas commonly arise from the U and Th chains and also from $^{40}$K and $^{60}$Co.  Second, electrons from beta decay isotopes, such as tritium, may have arbitrarily small energies and therefore can produce electron recoils in the signal region.  Nuclear recoils from decays at the detector surface, in which the electron is undetected, may also deposit small amounts of energy.

\item {\it Compton scattering:} Photons from radioactivity and atomic transitions in the detector material or housing may Compton scatter once at a shallow angle in the detector.  Since there is no discrimination between electronic and nuclear recoils, such shallow scattering would be indistinguishable from the neutrino signal.  While the rate of these events is obviously dependent on the level of radioactive contamination, we expect kinematics to strongly suppress the rate of these events.

\item {\it Photoelectrons:} Photons produced in the detector or housing may produce photoelectrons in the detector material, which could be ejected.  Recoils from such events could produce small energy depositions in the energy region of interest.  Low-energy secondaries from high-energy gammas produced in the detector or housing may impinge on other inactive material in the experiment and eject low-energy photoelectrons that could strike a detector.

\item {\it Photons from atomic relaxation transitions from the surrounding surfaces:} Photons from atomic transitions are of roughly the correct energy to produce some background events near a 10~eV threshold in a Si detector.  Copper, a good material for the detector housing for example, has 250 atomic lines with energies in the range 10-50~eV.  The rate expected from such events is very difficult to quantify, and would depend on the amount of low-energy radiation present in the cryostat to excite these transitions.  

\item {\it Neutrons:} Conventional methods used to reduce and model the neutron flux in dark matter experiments can be used.  Running the experiment at large overburden, a muon veto can be used to veto cosmogenic neutrons with high efficiency.  The background rate of neutrons from muons that miss the veto can also be estimated with simulation.  CDMS, for example, is able to achieve an unvetoed neutron rate of $< 0.1$ events/kg/year in the energy range 10keV to 100keV \cite{CDMS}.  Since the cross section for elastic scattering of neutrons on Si is fairly constant down to low energies, we do not expect the neutron background to be significant in our energy region of interest.

\item {\it Neutrino-electron scattering:} In addition to scattering coherently off nuclei, the neutrinos will also scatter off electrons in the detector material.  The cross section for this process is lower than the cross section for coherent neutrino scattering, and the recoil spectrum extends up to hundreds of keV for $^{37}$Ar neutrinos.  Since this background is also well-predicted by the standard model, we expect this to be a small contribution that may be reliably subtracted.

\item {\it Dark matter:} Recent experiments have reported signals that are consistent with a light WIMP of mass $\sim 7$~GeV and spin-independent cross section of $\sim 10^{-40}$~cm$^2$ for elastic scattering \cite{CoGeNT:2010,CoGeNT:2011}.  If the dark matter interpretation of these signals is correct, then scattering of WIMPs in the bolometers could be significant in the energy range of interest for coherent neutrino scattering.  With standard assumptions regarding the WIMP halo velocity, we estimate a rate of 0.16 events/kg/day in 10-50~eV for such a $\sim 7$~GeV WIMP.  Although this is a substantial rate, it is much smaller than our expected event rate and would be independent of the distance of each detector from the neutrino source.  Given the uncertainty surrounding the measurements in \cite{CoGeNT:2010,CoGeNT:2011} and the conflicting result reported in \cite{CDMS_LowEnergy}, we do not consider this background further in our analysis.
\end{itemize}

Excluding dark matter and the unknown backgrounds due to atomic transitions, the Compton scattering and photoelectron backgrounds are expected to dominate.  Using the raw rate of events in CDMS, we can conservatively estimate the background rate due to these two sources.  A good detector (250~g) in CDMS sees a raw rate of 0.001~Hz between 1--200~keV.  If we conservatively assume that these events all lie in the range 1--10~keV and furthermore are attributable only to Compton scattering and photoelectric effect, then the rate of events is 38.4 events/kg/day/keV, assuming a flat spectrum.  Assuming that the spectrum is flat down to 10~eV, we would see 1.54~events/kg/day in the energy range 10--50~eV.  In reality, we do not expect the spectrum to be flat down to low energies. Low-energy gammas from k- and l-shell electron captures are much more likely to be absorbed by the photoelectric effect, causing their full energy to be measured in the detector, and suppressing their background at lower energies. Only a small fraction of these produce ejected photoelectrons which could leave small energy depositions in the detector as described above. Compton scattering is very suppressed in the 1--10~keV range, and only dominates above 60~keV. The fraction of the background seen in CDMS detectors that lies above 60~keV is very small. For these reasons, the figure of 1.54~events/kg/day between 10--50~eV is a very conservative estimate of the Compton and photoelectron background.  For the sake of simplicity, we assume a 1 events/kg/day background in our energy window of interest.

\subsection{Detector Calibration}
Detector-to-detector variations constitute the most serious uncertainty in this measurement.  Differences in the yield at different distance could quickly mask --or worse yield a false-positive on-- the oscillation signal.  Part of the problem can be solved by calibrating the efficiency of the detectors using a neutron calibration source.  An attractive candidate would be to use a \isotope{H}{3}(\isotope{H}{3},2n)\isotope{He}{4} compact di-neutron source, such as used in the petroleum industry~\cite{bib:Schulumberger}.  The neutrons produced from this source should yield a relatively flat neutron energy spectrum, which is ideal for studying detector acceptance and to verify the $1/r^2$ response of the array.  

As shown in Figure~\ref{fig:array}, our approach is to remove the dependency of the oscillation measurement on the response of a particular detector.  The source is placed on a movable platform, and moved along the Si array throughout the measurement. Over the course of the experiment, each detector samples multiple baselines, and can be cross-calibrated with other detectors at each baseline to remove the individual detector response differences. Detector variations are essentially constrained by the in-situ measurement. 

It will also be important to calibrate the detector response to low energy photons and electrons.  Recent advances in solid state UV diodes make well-tuned eV photon sources readily available.  UV diodes ranging from 255 nm to 350 nm (3.5 eV - 5.2 eV) with sub-eV resolution are now commercially available.

\section{Sensitivity and Outlook}
\label{sec:summary}

Having discussed in detail both the source and the detection mechanism, we can now examine the signal in such an apparatus.  For a monochromatic and isotropic source with activity $R(t)$ encapsulated in some volume $V_S$, the signal rate as a function of time $t$ is given by the expression:

\begin{widetext}
\begin{equation}
\label{eq:signal}
S(t) = \sum_i R(t) \cdot \sigma_0(E_\nu) \cdot f(E_\nu,T_0) \cdot  \frac{N_A}{A} \cdot \rho_t \int  \frac{dV_s}{V_s} \int \frac{P(E_\nu, r_{st})}{4\pi r_{st}^2}~dV_{T,i}
\end{equation}
\end{widetext}

\noindent where $N_A$ is Avogadro's number, $\rho_t$ is the target density, $dV_{T,i}$ is the differential target volume of a single detector, and $r_{st}$ is the source-target distance.  The sum is taken over all discrete detectors available for the measurement.  In the approximation of a point source, Equation~\ref{eq:signal} reverts to the more familiar form:

\begin{equation}
\label{eq:signalsimple}
S(t) = \sum_i R(t) \cdot  \sigma_0(E_\nu) \cdot f(E_\nu,T_0) \cdot \frac{N_A}{A} \cdot M_t \cdot \frac{ P(E_\nu,\bar{r}_{i})}{4\pi \bar{r}_{i}^2}
\end{equation}

\noindent where $\bar{r}_i$ now is the average-weighted distance from the source to the individual detectors and $M_t$ is the mass of each detector. In the limit where the measurement time is much greater than the source half-life, the total number of accumulated signal events is given by $N \simeq S(t_0) \cdot \tau_{1\over2}/\ln{2}$, where $\tau$ is the half-life of the neutrino source. Extending measurements well beyond the peak source activity has the added benefit of reducing the statistical uncertainty on the background. 

For a monoenergetic source, the oscillation signal is all encoded within the spatial distribution of events.  A deviation from the expected $r^{-2}$ dependence could constitute a possible oscillation signal.  For the case where there is only one additional neutrino, the oscillation probability is given by the neutrino oscillation formula:
\begin{equation}
\label{eq:oscillations}
P(E_\nu, r) = 1 - \sin^2{(2\theta_S)} \sin^2{(1.27 \Delta m_S^2 \frac{r}{E_\nu})} 
\end{equation}
\noindent where $\sin^2{(2\theta_S)}$ is the amplitude to oscillate to the sterile state, and $\Delta m_S^2$ represents the sterile mass splitting.  In this case, $E_\nu$ is measured in units of MeV, $r$ in meters, and $\Delta m_S^2$ in eV$^2$.  For simplicity, we look at the simple 3+1 neutrino model, where the oscillation is to just one additional sterile neutrino. 

We use simulated data from a mock experiment to determine the potential sensitivity to sterile neutrinos.  We consider a compact 5 MCi \isotope{Ar}{37} source to be used in conjunction with a 500~kg silicon array.  We consider a total exposure of 300 days in order to extract both signal and background rates.  Parameters relevant for the fit are listed in Table~\ref{tab:values}.  For comparison, we also list the parameters for a germanium array with similar number of detectors and energy threshold. Due to the lower mass per detector needed to achieve the lower threshold and the lower recoil energies of neutrinos off the heavier germanium nucleus, a germanium array will achieve a signal rate that is about half of the silicon array. 

\begin{table}[htdp]
\caption{List of relevant source and detector parameters used for sensitivity analysis.  The signal rate is quoted for a {\em single} detector located 10 cm away from the center of a 5 MCi (185 PBq) \isotope{Ar}{37} source.}
\begin{center}
\begin{tabular}{|l|c c|}
\hline
Parameter &  \multicolumn{2}{|c|}{Detector Type} \\
\hline
Detector Material & Si & Ge \\
Atomic Number & 28 & 72.6 \\
$\sigma_0(E_\nu)$ $(10^{-42}$ cm$^2$) & 0.44 & 3.82\\
$T_{\rm max}$ & 50.3 eV& 19.4 eV\\
Threshold &  \multicolumn{2}{|c|}{10 eV}\\
$f(E_\nu,T_0)$ (see Eq. (\ref{eqn:fracAboveThresh})) & 64.2\% & 23.6\% \\
Detector cube size &  28 mm & 15.5 mm \\
Detector Mass & 50 g & 20 g\\
Number of Detectors &  \multicolumn{2}{|c|}{10,000} \\
Total Mass & 500 kg & 200 kg \\
Yield at 10 cm (kg$^{-1}$day$^{-1}$MCi$^{-1}$) & 15.28& 19.0 \\
Signal Rate at 10 cm & 3.82 day$^{-1}$ & 1.90 day$^{-1}$ \\
\hline
\end{tabular}
\end{center}
\label{tab:values}
\end{table}%

\begin{figure}[htbp]
\begin{center}
\includegraphics[width=0.9\columnwidth]{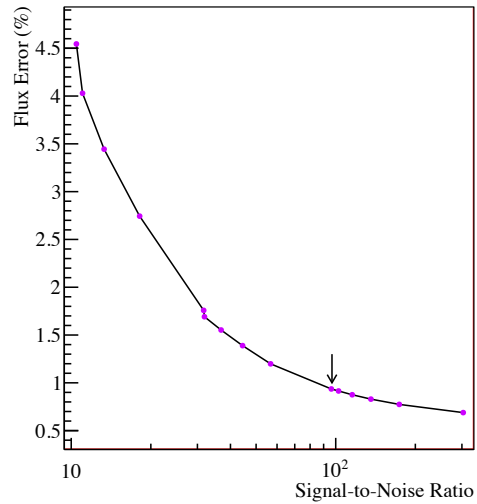}
\caption{Plot of the relative signal error versus signal-to-noise ratio $\frac{S}{\sqrt{B}}$ ($S$ represents signal strength, $B$ represents background counts) for a 500-kg Si array exposed to a 5 MCi \isotope{Ar}{37} for 300 days.  This array configuration and source intensity yields approximately $S \simeq 54,000$ total signal events.  Arrow indicates signal-to-noise ratio corresponding to 1 background event/kg/day.}
\label{fig:backgrounds}
\end{center}
\end{figure}

For such an experiment, we also consider a number of systematic errors:
\begin{itemize}
\item {\it Source Strength:}  The SAGE collaboration used a variety of techniques in order to determine the final \isotope{Ar}{37} activity, including gas volume, gas mass, calorimetry, direct counting and isotopic dilution.  Any one of these methods in isolation achieved a $\pm 1\%$ accuracy, while in conjunction the total uncertainty was less than $\pm 0.5\%$.  In this study, we assume a conservative $\pm 1\%$ uncertainty on the source strength.  Since the source uncertainty applies to all detectors globally, it has minimal impact on the oscillometry measurement.
\item {\it Cross-section:}  The cross-section uncertainty, much like the source strength uncertainty, is a global uncertainty and has little impact on our oscillometry extraction.  Its uncertainty would nominally be dominated by the uncertainty in the form factor, but at such exchange momenta the effect is expected to be small.  We therefore assume a $\pm 1\%$ global uncertainty due to the cross-section.

\item{\it Vertex Resolution:} The bolometric detector in this experimental design is composed of 10,000 silicon or germanium absorbers instrumented with a single thermometer. The dimensions of these absorber cubes are 28 and 15.5~mm per side for silicon and germanium, respectively. These dimensions are smaller than the source itself (assumed to have a radius of 4 cm), thus the vertex resolution is dominated solely by the extension of the source.  This effect is incorporated into our analysis.

\item {\it Detector Variations:}  Detector variations are kept under control via the series of in-situ and ex-situ calibration measurements discussed in the previous section.  Using the movable source depicted in Fig.~\ref{fig:array}, one should be able to calibrate the detector variations to about $\pm 2\%$.  The global uncertainty, which also depends on fiducial volume dependence, overall efficiency, etc., is estimated to be $\pm 5\%$.

\item {\it Detector Backgrounds:}  With detector-to-detector variations calibrated away, the main challenge for such a measurement remains the number of detector backgrounds that accumulate during the measurement.  As discussed above, the question of what the background will be between 10--50~eV is hard to estimate at this point, and more work needs to be done to enable a credible estimate. For this study, we assume a total background activity of 1 event/kg/day in the signal region of interest.  The signal-to-noise ratio should scale roughly as the square root of the number of background events.  The dependence of the accuracy of the measurement as a function of then signal-to-noise ratio is shown in Fig.~\ref{fig:backgrounds}.  Measurements taken with background levels below 1 event/kg/day are essentially systematics dominated.

\item {\it Source-Induced Backgrounds:}  Any backgrounds that stem directly from the \isotope{Ar}{37} source may potentially dilute the sensitivity of the measurement, since they, too, would exhibit a $1/r^2$ behavior.  The SAGE group has produced an extremely pure argon source, with less than 1\% of the volume having \isotope{Ar}{39} contamination.  Given the extremely long half-life of \isotope{Ar}{39} and the vastly different signature ($\beta$-decay), we believe this is a negligible background source.  Though the majority of the energy from the decay of \isotope{Ar}{37} is removed by neutrinos, a fraction of the energy is carried away from recoils and internal-bremsstrahlung photons.  The SAGE source effectively reduced this contribution to less than $0.2\%$.  Effective shielding should reduce this contribution even further.

\item{\it Other Neutrino Interactions:}  One of the isotopes of germanium (\isotope{Ge}{71}) has a low enough threshold to allow $\nu_e$ charged current scattering. However, with a threshold energy of 321 keV, the outgoing electron will have recoil energy far above the energy region of interest.  Hence, its contribution to the overall background can be considered negligible.  Charged current interactions on silicon all have thresholds above 1 MeV, hence they do not contribute to the background activity.  As discussed previously, other charged-current interactions, such $\nu_e e^-$ elastic scattering, are highly suppressed. As such, their contribution is also expected to be negligible.

\end{itemize}

\begin{table}[htdp]
\caption{List of systematic uncertainties expected for a low-threshold germanium detector array.  Uncertainties are listed for both shape+rate and shape only analysis.}
\begin{center}
\begin{tabular}{|l|c|c|}
\hline
Source & \multicolumn{2}{|c|}{Systematic} \\
\hline
& ~Global~ & Shape Only \\
\hline
Source Strength & $\pm 1\%$ & - \\
Cross-section & $\pm 1\%$  & - \\
Detector Variation& $\pm 2\%$ & $\pm 2\%$ \\
Absolute Efficiency& $\pm 5\%$ & - \\
Source-Induced Background & $< 1\%$ & $< 1\%$ \\
Vertex Resolution & $\pm 2.8$ cm & $\pm 2.8$ cm\\
Source Extent & $\pm 4$ cm & $\pm 4$ cm \\
\hline
Total Systematic & $\pm 5.5\%$ & $\pm 2\%$\\
\hline
Statistical (Whole Array) &  \multicolumn{2}{|c|}{$\pm 1\%$}\\
\hline
\end{tabular}
\end{center}
\label{tab:sys}
\end{table}%

\begin{figure}[htbp]
\begin{center}
\includegraphics[width=0.9\columnwidth,keepaspectratio=true]{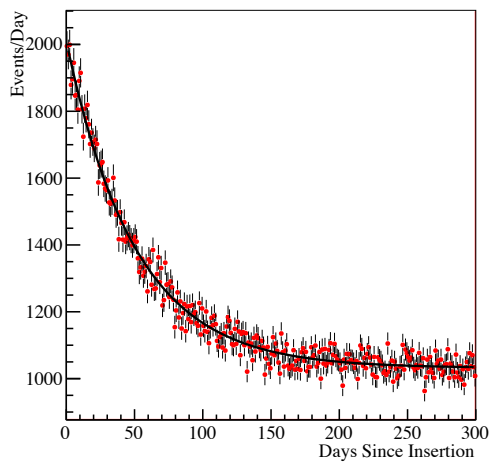} 
\caption{Distribution of events on a 500 kg Si array as a function of time of source deployment.  Source considered here is a 5 MCi \isotope{Ar}{37} electron-capture source.}
\label{fig:time}
\end{center}
\end{figure}

A summary of the relevant systematic uncertainties are listed in Table~\ref{tab:sys}. A simple $\chi^2$-fit is used to estimate the sensitivity of the proposed Si and Ge arrays.  The data extracted from the entire array is first fit as a function of time in order to extract the overall source strength and background (see, for example, Figure~\ref{fig:time}).  The background-subtracted signal is then fit to the oscillation formula of Eq.~\ref{eq:oscillations}. The analysis uses both shape and rate to determine the sensitivity to sterile neutrinos. In the case of the shape+rate analysis, an additional penalty term is added to the likelihood from the overall flux measurement.  

\begin{figure*}[htbp]
\begin{center}
\subfigure{\label{fig:length}\includegraphics[width=0.9\columnwidth]{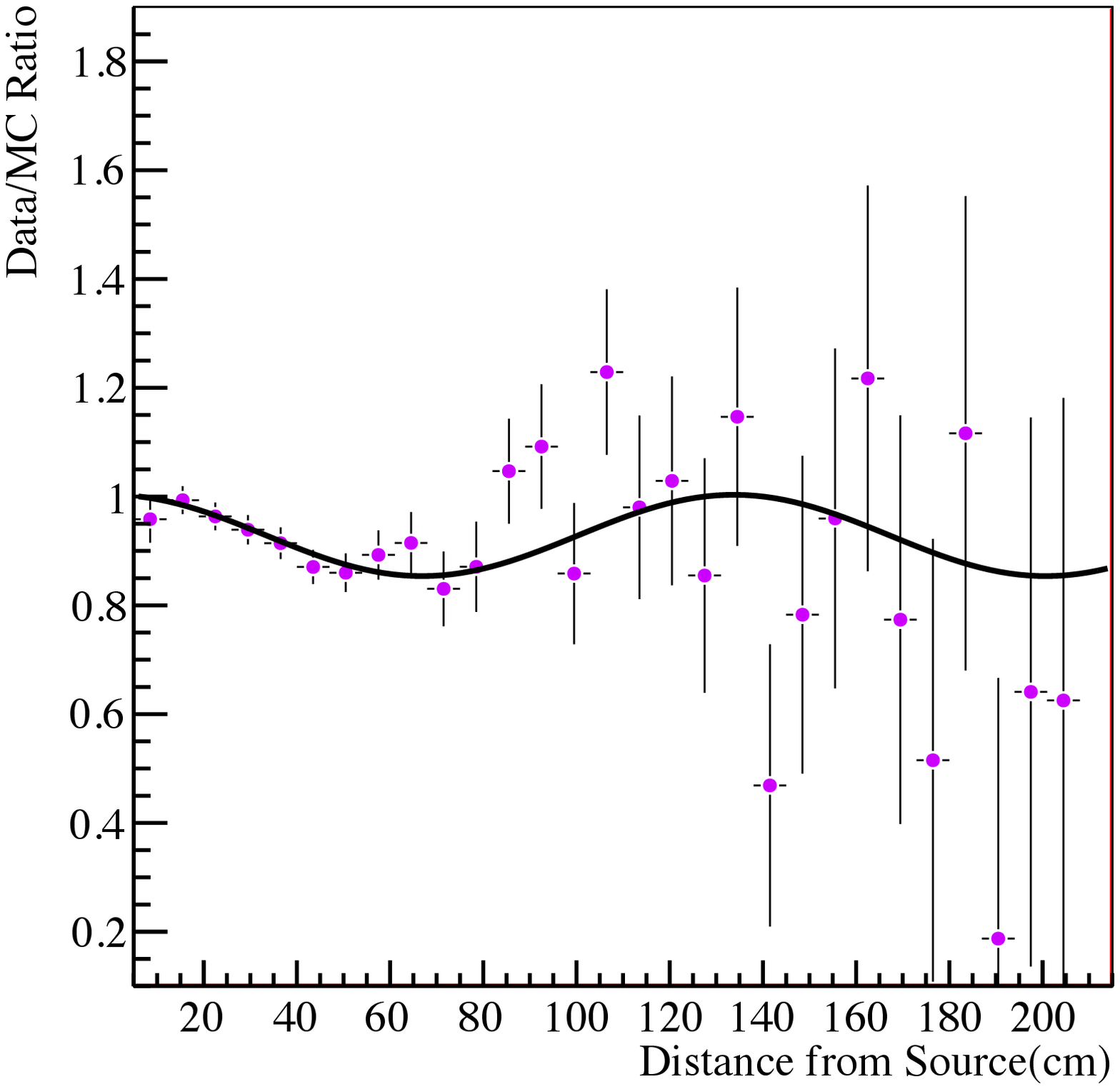}}
\subfigure{\label{fig:thesignal}\includegraphics[width=0.9\columnwidth]{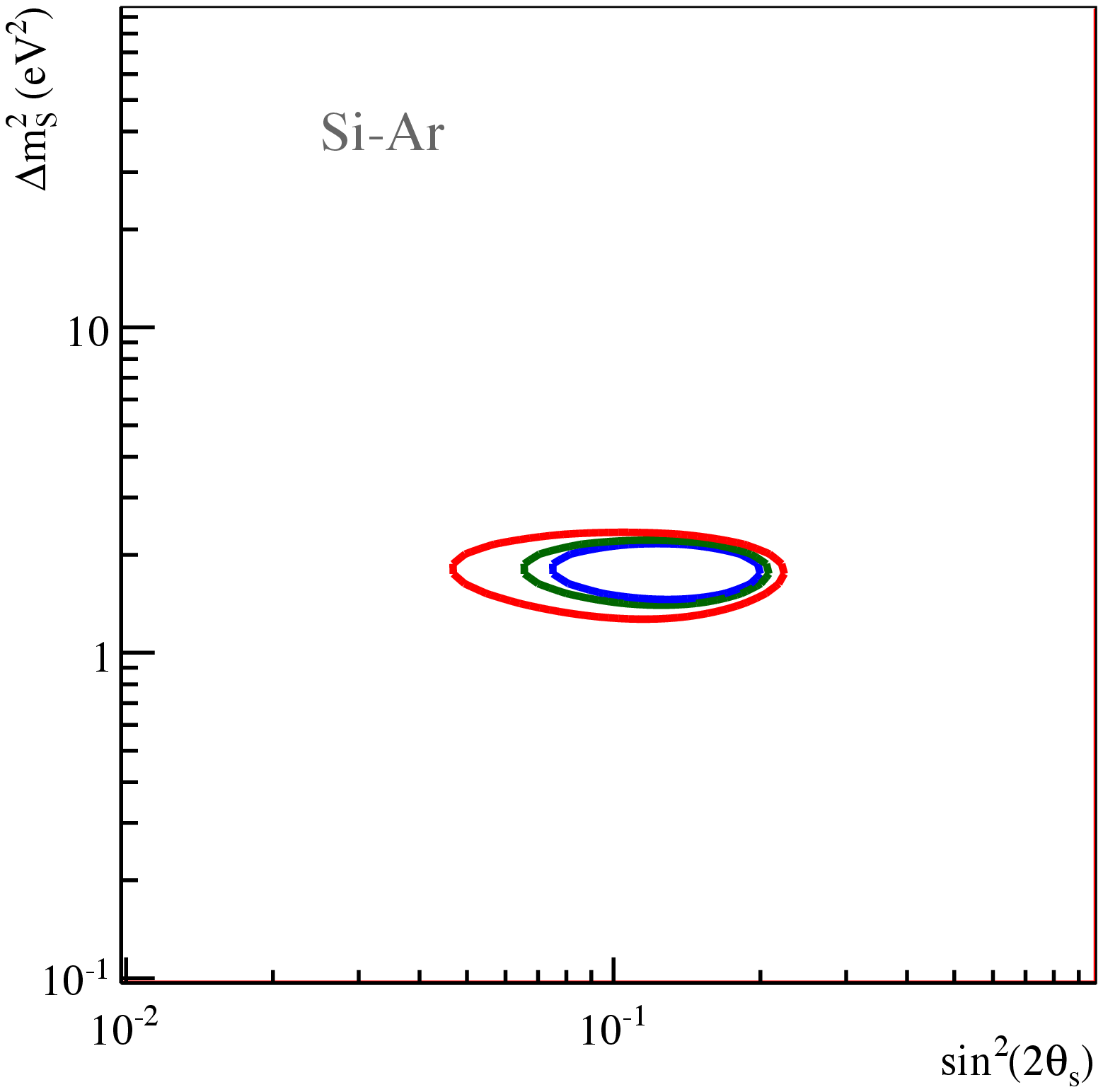}}
\caption{Left: Ratio of data and Monte Carlo for a simulated neutrino oscillation signal ($\Delta m_S^2$ = 1.5 eV$^2$, $\sin{(2\theta_S)^2} = 0.15$) as a function of source distance from a 5 MCi \isotope{Ar}{37} neutrino source and a 500 kg Si-array. Right: Likelihood contour curves for same signal after 300 days of data taking.  Contour levels are shown at 90\% (blue), 95\% (green), and 99\% (red).}
\label{fig:results}
\end{center}
\end{figure*}

\begin{figure*}[htbp]
\begin{center}
\begin{tabular}{c c}
\subfigure[~Silicon - \isotope{Ar}{37}]{\label{fig:resultsSiAr}\includegraphics[width=0.95\columnwidth]{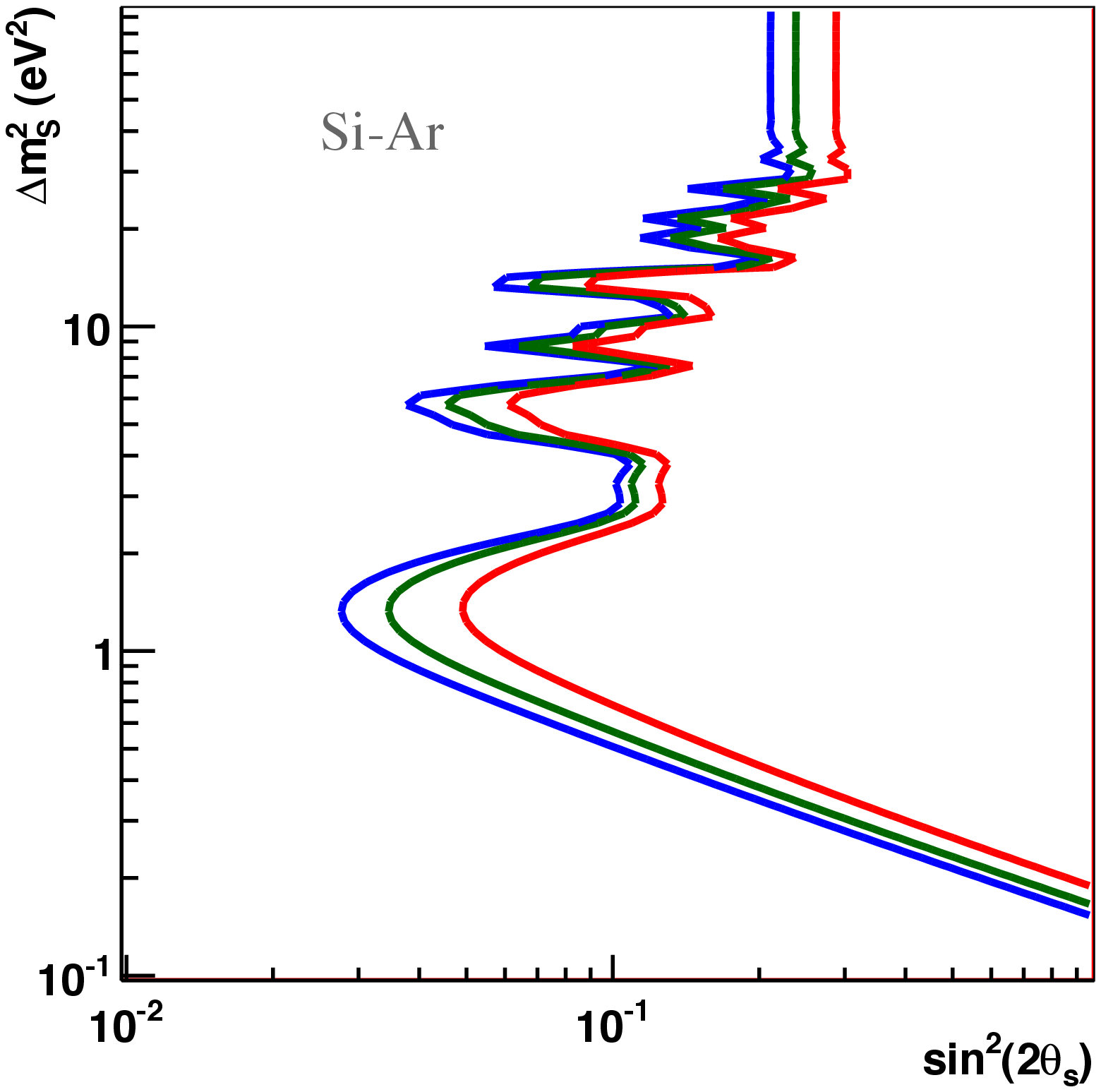}} &
\subfigure[~Germanium - \isotope{Cr}{51}]{\label{fig:resultsGeAr}\includegraphics[width=0.95\columnwidth]{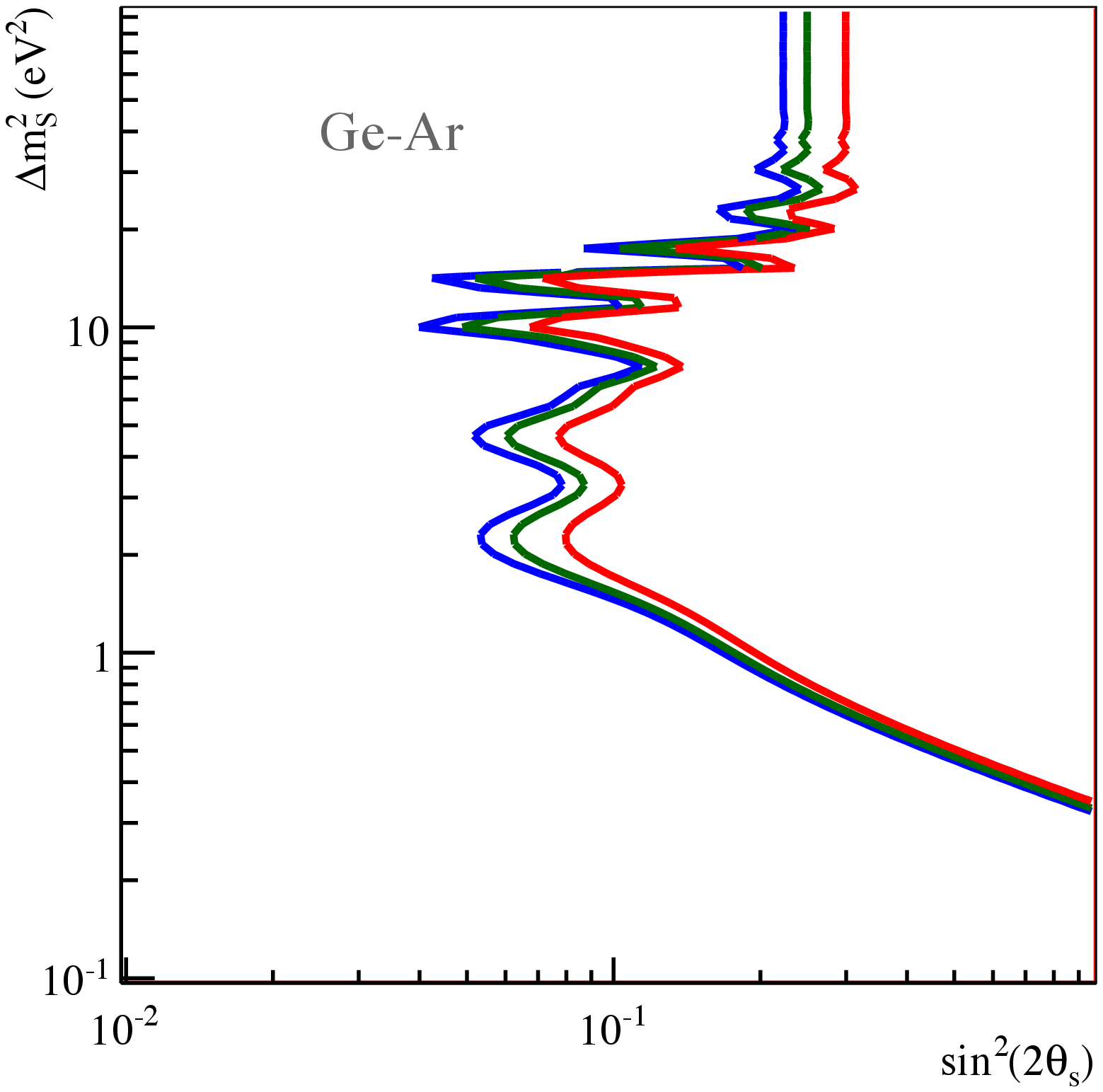}} \\
\subfigure[~Silicon - \isotope{Ar}{37}]{\label{fig:resultsSiCr}\includegraphics[width=0.95\columnwidth]{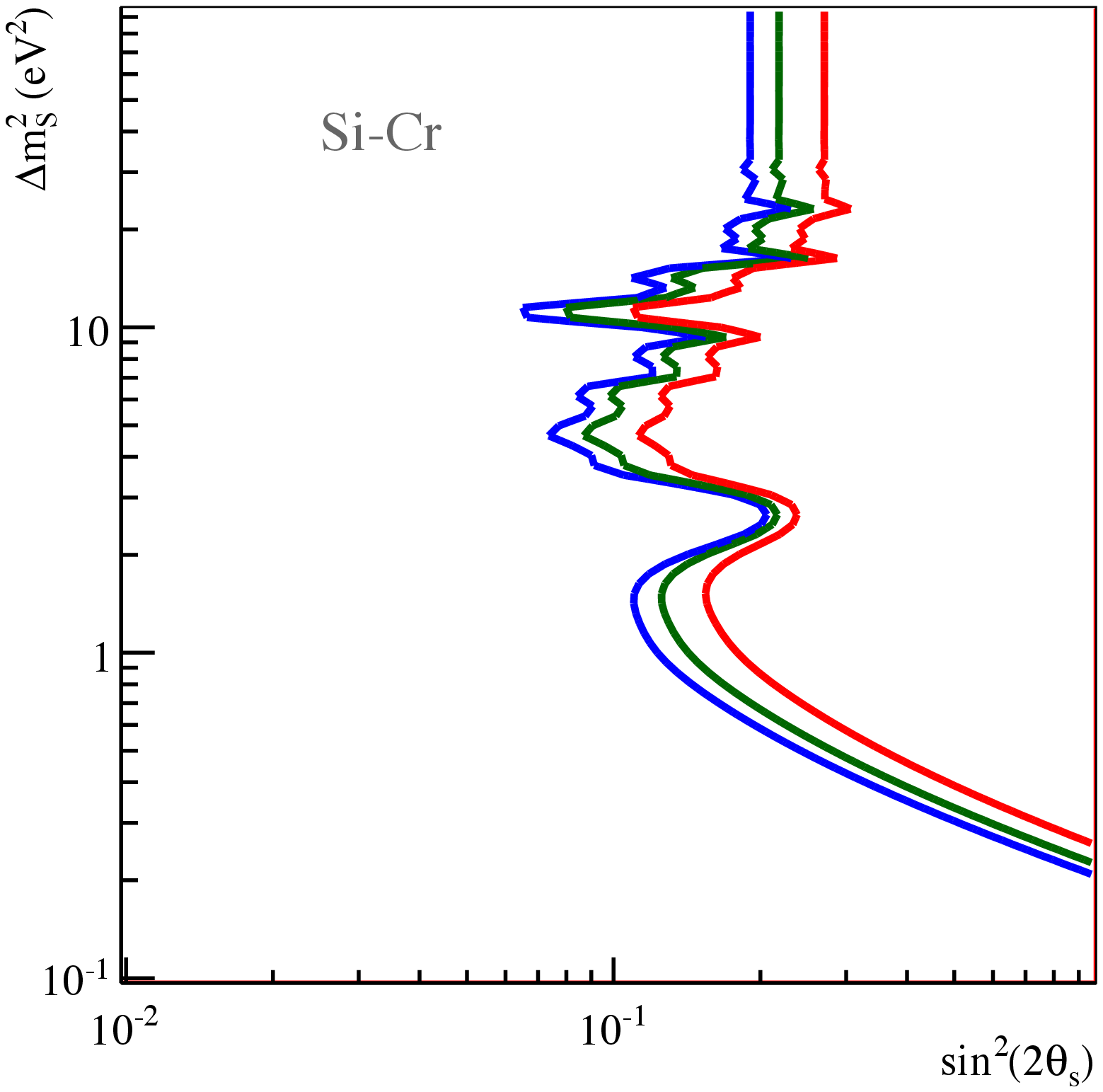}} &
\subfigure[~Germanium - \isotope{Cr}{51}]{\label{fig:resultsGeCr}\includegraphics[width=0.95\columnwidth]{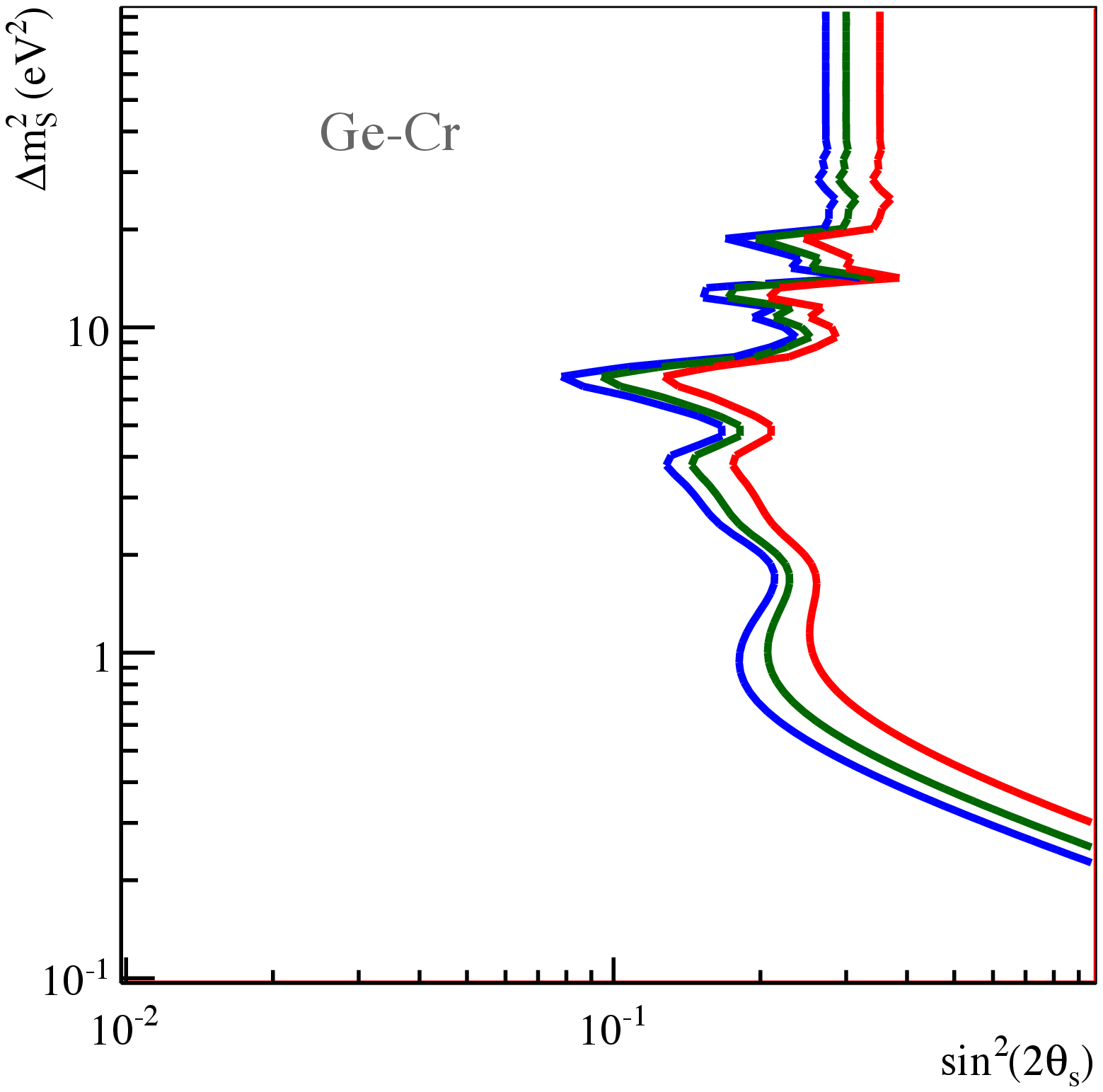}} \\
\end{tabular}
\caption{Likelihood contours for a 300-day run on a 500 kg Si array (left) and 200 kg Ge (right) array exposed to a 5 MCi \isotope{Ar}{37} (top) and \isotope{Cr}{51} (bottom) source, using both shape and rate information. Confidence levels in all plots are shown at 90\% (blue), 95\% (green), and 99\% (red). Statistical and systematic errors are included in the signal analysis.}
\end{center}
\end{figure*}

\begin{figure}[htbp]
\begin{center}
\includegraphics[width=0.9\columnwidth]{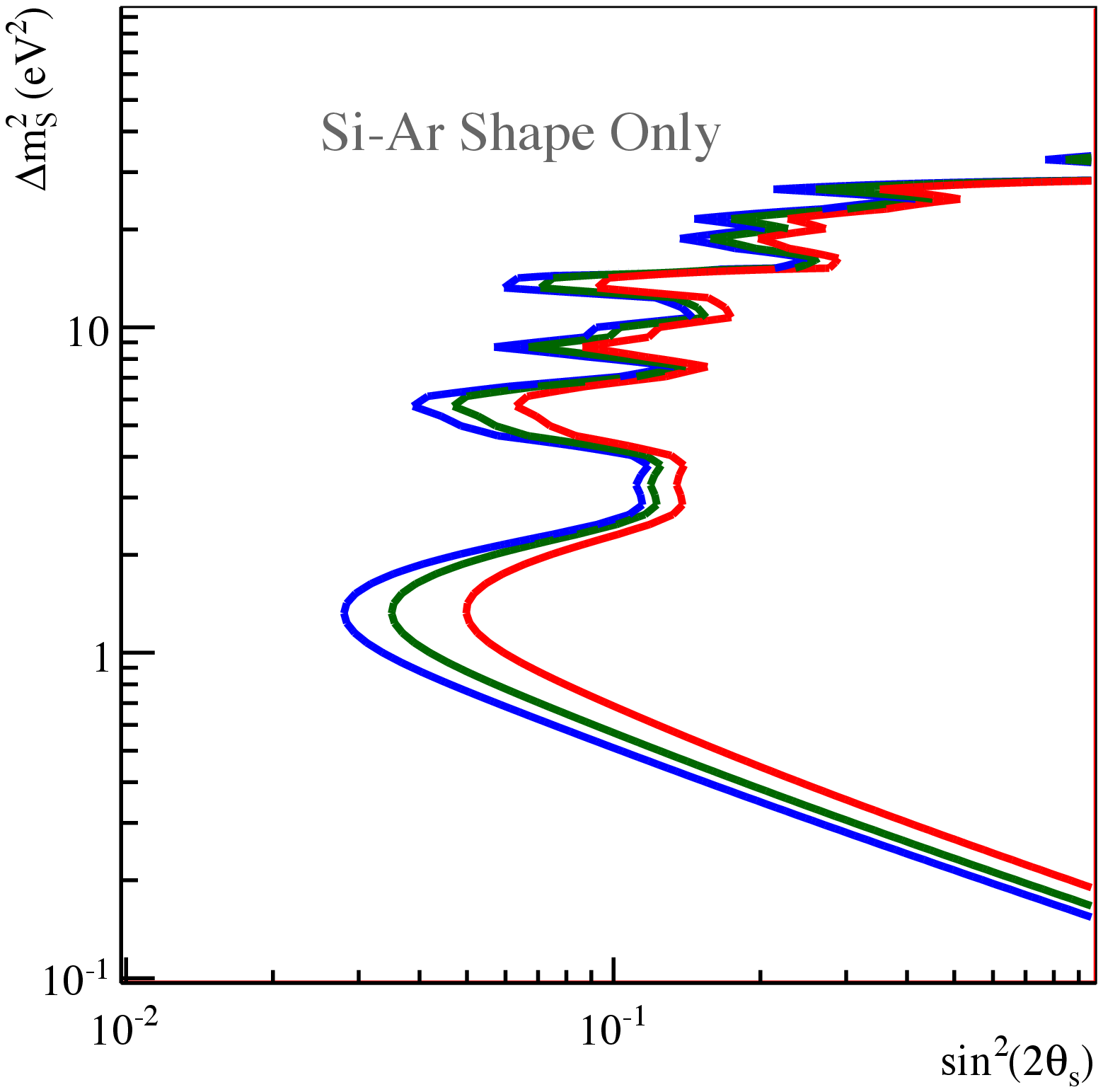}
\caption{Likelihood contours for a 300-day run on a 500 kg Si array and a 5 MCi \isotope{Ar}{37} source, using only shape information. Contour curves use same scheme as in Figure~\ref{fig:results}.}
\label{fig:shapeonly}
\end{center}
\end{figure}

Results for a 500 kg Si detector array are shown in Fig.~\ref{fig:resultsSiAr}. The distortion caused by a non-zero sterile mixing is statistically distinguishable in the measured distance profile (see Figure~\ref{fig:length}).  As can also be seen from the figure, the array is not necessarily fully optimized for a given oscillation length scale.  Such optimization can proceed once the parameter space for sterile neutrinos is further constrained by ongoing and future neutrino experiments.  Nevertheless, for the bulk of the region of $\Delta m_S^2 = 1-10$ eV$^2$ and $\sin{(2\theta_S)}^2 \ge 0.08$, typically preferred from the reactor data, is ruled out at the 90\% C.L.  If the best fit solution from the reactor anomaly is viable, then the measurement should be detectable at the 99\% C.L. (see Fig.~\ref{fig:thesignal}). It is possible to also conduct a shape-only analysis.  Most of the sensitivity to sterile oscillations is retained for $\Delta m_S^2$ masses below 10 eV$^2$.  A sensitivity curve for a shape only analysis is shown in Figure~\ref{fig:shapeonly}.  

For comparison, we also consider an equivalent Ge array with a total mass of 200 kg.  These results are shown in Fig.~\ref{fig:resultsGeAr}.  Finally, for completeness we also show the detector sensitivity for the Ge and Si arrays using an equivalent \isotope{Cr}{51} radioactive neutrino source (Figures~\ref{fig:resultsSiCr} and~\ref{fig:resultsGeCr}). In general, the reduced source energy decreases the available statistics, so a relatively stronger source needs to be considered in such a case.

\section{Summary}

We have outlined the possibility of probing the existence of sterile neutrinos using coherent scattering on a bolometric array.  Such a method could provide the most direct test of oscillations to sterile neutrinos.  With the advent of low threshold detectors and the use of intense neutrino sources, such an experiment appears feasible with our current technology.  Such a program is also very complimentary to any existing dark matter search.

Even in the absence of sterile neutrinos, the experiment as described in this letter can make other important measurements.  Most prominently, such an experiment may constitute the first observation of coherent scattering.  For a 500 kg detector, it should be able to make a $\simeq 5\%$ measurement on the overall cross-section, pending on the absolute calibration of the efficiency.  For an isoscalar target, such as silicon, this provides a direct measurement of the weak mixing angle at momentum transfer as low as 1 MeV.  


\section{Acknowledgments}
\label{sec:Acknowledgments}

The authors would like to thank Bruce Cleveland for his insight on the production of neutrino calibration sources, Scott Hertel, who was the catalyst for bringing these ideas together, Simon Bandler, George Seidel, Blas Cabrera and Steve Leman for useful discussions on heat capacities, Rupak Mahapatra and Rusty Harris for useful discussions on backgrounds and detector fabrication, and Jack Sadleir for discussions on TES properties at 15~mK.  J. A. Formaggio is supported by the United States Department of Energy under Grant No. DE-FG02-06ER-41420, and E. Figueroa-Feliciano is supported by the United States National Science Foundation under Grant No. PHY-0847342.

\bibliographystyle{apsrev}
\bibliography{CoherentSterile}

\end{document}